\input{aipcheck}
\documentclass[english,final]{aipproc}
\layoutstyle{6x9}
\usepackage{graphicx}
\usepackage{epsfig}
\usepackage{xspace}
\usepackage{amsmath}
\usepackage{amssymb}
\usepackage{babel}

\begin{document}

\title{The Double Chooz Experiment\footnote{To appear in Proceedings of the Conference on the Intersections of Particle and Nuclear Physics, Puerto Rico, May 30 - June 3, 2006.}}



\author{Daniel M. Kaplan (for the Double Chooz Collaboration)}{address=
{Illinois Institute of Technology,
3101 South Dearborn Street,
Chicago, Illinois, USA}}
 \begin{abstract} 
There is broad consensus in the worldwide physics community as to the need for a new reactor-neutrino experiment to measure or limit the neutrino mixing angle $\theta_{13}$. The Double Chooz Experiment, planned for operation in the years 2008$-$2011, 
will search for values of $\sin^2{2\theta_{13}}$ down to $\approx$0.03. This will be the first new information on $\theta_{13}$ in over a decade and will cover most of the remaining parameter space.
A quick and relatively inexpensive project is made possible by the existing neutrino laboratory at the Chooz site. 
\end{abstract}
 
\maketitle
\section{Introduction}
There is consensus within the physics community that an experiment should be mounted to measure the disappearance of reactor antineutrinos with sensitivity to the $\theta_{13}$ neutrino mixing angle an order of magnitude beyond that of previous experiments~\cite{APS-study,whitepaper}. Such an experiment (known as Double Chooz) has been proposed by an international collaboration~\cite{collab}, to be constructed using the existing neutrino laboratory at the Chooz power station in France~\cite{DoubleChooz}, and the proposal has been approved and funded in Europe. 

The Double Chooz experiment is more modest in scope (and in cost) than other proposed reactor-neutrino efforts~\cite{reactor}, aiming at sensitivity to $\sin^2{2\theta_{13}}\approx0.03,$\footnotemark
\addtocounter{footnote}{-1} while the goal of the other proposals is 0.01. Since it represents a more limited extrapolation from current capabilities, there is less need for fundamental R\&D to prove feasibility.
Moreover, the laboratory from a previous neutrino experiment done at Chooz~\cite{CHOOZ} is available for our use. Double Chooz can thus be put into operation quickly and with a well-known background environment. We expect that Double Chooz will provide important experience applicable to a more precise measurement of $\theta_{13}$ in future experiments. 

\section{Physics Motivation}

Of the three mixing angles characterizing three-flavor neutrino oscillation, two ($\theta_{12}$ and $\theta_{23}$) have been measured to be  large, but for the third ($\theta_{13}$), as yet there are only upper limits~\cite{PDG}, of which the most stringent ($\sin^2{2\theta_{13}}<0.16$ at 90\% CL\footnote{Sensitivity to $\theta_{13}$ depends on the assumed value of $\Delta m_{13}^2$. The benchmark value $\Delta m_{13}^2=2.5\times10^{-3}$\,eV$^2$ is used throughout this paper.}) comes from the previous Chooz experiment~\cite{CHOOZ}. Double~Chooz~\cite{DoubleChooz} is a reactor-neutrino disappearance experiment that can measure the $\theta_{13}$ neutrino mixing angle over most of the remaining allowed parameter space. 
Such an experiment is complementary to long-baseline accelerator experiments~\cite{reactor-complementarity,whitepaper}, since it avoids ambiguities from {\em CP} violation and matter effects. 
Because neutrino
{\em CP} violation can be observed only when the initial and final states differ, a disappearance experiment is insensitive to the phase $\delta$ of the PMNS neutrino mixing matrix;  the low neutrino energies and short baselines employed eliminate sensitivity to matter effects. A reactor-neutrino experiment thus makes a clean measurement of $\sin^2{2\theta_{13}}$.
Double~Chooz will also provide crucial  guidance to future accelerator long-baseline efforts (NO$\nu$A~\cite{NOVA} {\it et al.}) by limiting the range of parameter space over which they must be optimized, and to proposed, more ambitious, reactor experiments (Angra, Braidwood, Daya Bay, {\it et al.}~\cite{reactor}) by demonstrating  techniques they will need to use to control  
systematic errors, but in a less stringent regime.

\section{Experimental Approach}

\begin{figure}
\hspace{.15in}\scalebox{0.23}{\includegraphics{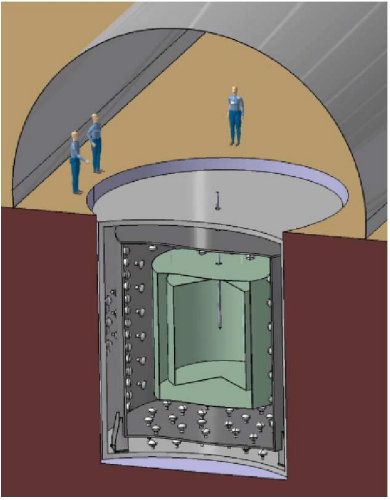}}\scalebox{.7}{\hspace{0.2in}\includegraphics{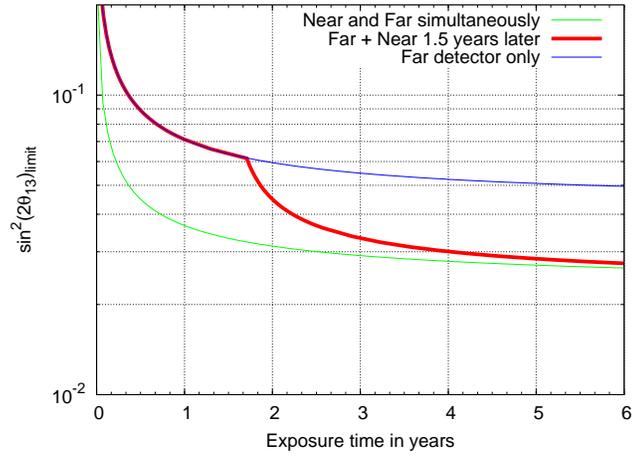}}
\caption{Left: cutaway drawing of a Double Chooz detector showing innermost (acrylic) target vessel surrounded in turn by (acrylic) $\gamma$-catcher vessel, (stainless) buffer/PMT-support vessel, and inner-veto vessel. Right: Double Chooz limit on $\sin^2{2\theta_{13}}$ vs.\ time (for $\Delta m_{13}^2=2.5\times10^{-3}$\,eV$^2$) assuming near-detector completion 18 months after far-detector turn-on.}
\label{fig}
\end{figure}

Like the first Chooz experiment~\cite{CHOOZ}, Double Chooz will employ Gd-loaded liquid scintillator as the neutrino detection medium. Gadolinium's large  neutron-capture cross section makes it the nucleus of choice for detection of the inverse-beta-decay reaction ${\overline \nu}_e+p\to e^++n$; the signature for antineutrino interaction is thus a pair of prompt 511\,keV gammas from positron annihilation (boosted by the $\sim$MeV positron kinetic energy) followed some tens of microseconds later by 8\,MeV of gammas from de-excitation of the Gd nucleus. Figure~\ref{fig} (left) illustrates the detector design: a series of concentric cylindrical tanks, of which the (transparent) innermost two are filled (respectively)  with Gd-loaded  (``target'') and unloaded (``$\gamma$-catcher'')  scintillator, and the (steel) outer two with mineral oil (``buffer'') and scintillator (``inner veto''). Photomultiplier tubes (PMTs) mounted on the interiors of the steel tanks detect the scintillation light  emitted in the target, $\gamma$-catcher, and inner veto; the nonscintillating buffer serves to suppress counting rate due to background radioactivity.  Outer veto detectors located above the nested tanks will provide additional cosmic-ray-muon rejection. Some key parameters of the experiment are listed in Table~\ref{tab:params}. The expected systematic errors in Double~Chooz are compared with those achieved in Chooz in Table~\ref{tab:Chooz-syst}. 
\begin{table}
\begin{tabular}{lrc}
\hline 
Parameter & Value\tablenote{Parameter values from Double Chooz proposal~\protect\cite{DoubleChooz}.} & Comment \\
\hline
Thermal power & 4.25 GW & each of 2 cores \\
Electric power & 1.5 GWe & each of 2 cores \\
${\overline \nu}_e$ target volume & 10.3 m$^3$ & Gd loaded LS (0.1\%) \\
$\gamma$-catcher thickness & 55 cm & Gd-free LS\\
Buffer thickness & 105~cm & nonscintillating oil\\
Total liquid volume & $\sim$237~m$^3$ & \\
Number \& size of phototubes per detector & 534 8{\tt "} & 13\% coverage \\
Far detector distance  & 1050~m & average\\
Near detector distance & 280~m & average\\
Far detector overburden & 300 m.w.e. & hill topology\\
Near detector overburden & 70$-$80 m.w.e. & shaft\\
${\overline \nu}_e$ far detector events (5 yr) & 75,000 & with 60.5\% efficiency\\
${\overline \nu}_e$ near detector events (5 yr) & 789,000 & with 43.7\% efficiency\\
Relative normalization error        & 0.5\% & \\
Effective bin-to-bin error       & 1\%   & background systematics \\
Running time with far detector only & 1$-$1.5 year & \\
Running time with far+near detector & 3 years & \\
$\sin^2{2\theta_{13}}$ goal in 3 years with 2 detectors & 0.02$-$0.03 & (90\% CL) \\
\hline
\end{tabular}
\caption{Main parameters of the Double~Chooz experiment.}
\label{tab:params}
\vspace{-.1in}
\end{table}

\begin{table}
\begin{tabular}{lcc}
\hline
 &\multicolumn{2}{c}{Relative error (\%)} \\
\raisebox{1.5ex}[0pt]{Error source} &Chooz & Double~Chooz \\
\hline
Reaction cross section & 1.9 & --- \\
Detection efficiency & 1.5 & 0.4 \\
Number of target protons & 0.8 & 0.2 \\
Reactor power & 0.7 & --- \\
Energy per fission & 0.6 & --- \\
\hline
Total & 2.7 & 0.5 \\ \hline
\end{tabular}
\caption{Comparison of Chooz~\protect\cite{CHOOZ} and Double~Chooz~\protect\cite{DoubleChooz} systematic-error contributions (in descending order of importance).}\label{tab:Chooz-syst}
\end{table}
Double~Chooz will improve upon Chooz in the following important respects:
\begin{enumerate}
\item {\em Near detector:} The addition of a near detector, of a size and detection technology identical to those of the far detector, will minimize uncertainties arising from the reactor neutrino flux and energy spectrum, 
neutrino cross sections, target volume, and detection efficiency. (The large signal rate at the near detector lessens the needed overburden, reducing costs.)
\item {\em Nonscintillating buffer:} A 1.05\,m thickness of nonscintillating liquid surrounding the neutrino target and $\gamma$-catcher will lower the singles rate by about two orders of magnitude compared to that in Chooz by suppressing counts due to irreducible sources of external radioactivity (dominated by trace radioactive isotopes in PMT constituent materials). The consequent reduction in accidentals rate will allow operation with a substantially lower ($\approx 500\,$keV) energy threshold, thereby reducing the systematic uncertainty in detection efficiency due to this threshold (a 0.8\% error contribution in Chooz). It will also allow backgrounds below 1\,MeV to be directly measured and improve the ability to intercalibrate the near- and far-detector energy scales.
\item {\em Scintillator stability:} The Chooz experiment experienced a gradual deterioration of light yield in the Gd-loaded scintillator  over time. While this degradation was not a dominant systematic error for Chooz, it is much more of a concern for Double Chooz. Consequently, the Double Chooz scintillator is being engineered for much better stability than that used in Chooz:  for samples of each formulation under study (Gd-beta-diketonate and Gd-carboxylate in a 20\%/80\% PXE/dodecane solution), one-year tests at room temperature show no appreciable decrease in light yield.
\item {\em Sample size:} The Chooz experiment operated during the startup period of the Chooz-B nuclear power station and accumulated a total of $2.13\times10^4$\,GWh of exposure during 1.3\,y of operation. With the power station now running stably at its full 8.5\,GW power, this exposure will be accumulated in Double~Chooz within the first 4 months; the total in 5\,y of running will exceed the Chooz exposure by a factor of $\approx$15. Along with a near-doubling of the target mass, this will yield some 75,000 far-detector ${\overline \nu}_e$ events and a 90\%-CL $\sin^2{2\theta_{13}}$ sensitivity of 0.03 at the benchmark value of $\Delta m^2_{31}$.
\end{enumerate}
Figure~\ref{fig} (right) illustrates the expected sensitivity vs.\ time, under the assumption that the near detector comes on-line 18 months after the far detector. The ability to keep systematics under 0.5\% will depend on careful calibration of the detectors and control of backgrounds, to which great attention is being paid. As of this writing, funding in the U.S. (requested from both DOE and NSF) is still under negotiation; nevertheless, the project is on schedule for a 2008 turn-on.

\end{document}